\documentclass[
,showpacs,amssymb,superscriptaddress,aps
]{revtex4}
\usepackage{graphicx}
\usepackage{color}
\input{epsf}

\usepackage{amsmath,amssymb}
\usepackage{bm}
\usepackage{times}
\usepackage{ulem}
\newcommand{\dalm}{\kern1pt\vbox{\hrule height 0.9pt\hbox{\vrule width 0.9pt
\hskip 2.5pt\vbox{\vskip 5.5pt}\hskip 3pt\vrule width 0.3pt}\hrule height 0.3pt}
\kern1pt}

\newcommand{\lsim}{\, \, \raisebox{-0.8ex}{$\stackrel{\textstyle <}{\sim}$ }}

\begin{document}

%\twocolumn[\hsize\textwidth\columnwidth\hsize\csname @twocolumnfals\endcsname

% For two column
%\wideabs{

\title{Estimating the nuclear saturation parameter via low-mass neutron star asteroseismology}
%\title{Imprint of the hyperon onset density in gravitational waves}

\author{Hajime Sotani}
\email{sotani@yukawa.kyoto-u.ac.jp}
\affiliation{Astrophysical Big Bang Laboratory, RIKEN, Saitama 351-0198, Japan}
\affiliation{Interdisciplinary Theoretical \& Mathematical Science Program (iTHEMS), RIKEN, Saitama 351-0198, Japan}

%\author{Hajime Togashi}
%\affiliation{}

\date{\today}

% Abstract
\begin{abstract}
We examine the fundamental ($f$-) and the 1st pressure ($p_1$-) mode frequencies in gravitational waves from cold neutron stars constructed with various unified realistic equations of state. With the calculated frequencies, we derive the empirical formulae for the $f$- and $p_1$-mode frequencies, $f_f$ and $f_{p_1}$, as a function of the square root of the stellar average density and the parameter ($\eta$), which is a combination of the nuclear saturation parameters. With our empirical formulae, we show that by simultaneously observing the $f$- and $p_1$-mode gravitational waves, when $1.5\lsim f_{p_1}/f_f\lsim 2.5$ (which corresponds to neutron star models with the mass of $\lesssim 0.9M_\odot$), one could estimate the value of $\eta$ within $\sim 10\%$ accuracy, which makes a strong constraint on the EOS for neutron star matter. In addition, we find that the maximum $f$-mode frequency is strongly associated with the minimum radius of neutron star. That is, if one would observe a larger frequency of the $f$-mode, one might constrain the upper limit of the minimum neutron star radius. 
\end{abstract}

%
%%%%%%%%%%%%%%%%%%%%%%%%%%%%%%%%%%%%%%%%%%%%%%%%%
%  04.30.-w  :  Gravitational waves
%  04.40.Dg :  Relativistic stars: structure, stability, and oscillations (see also 97.60.-s Late stages of stellar evolution) 
%  21.65.Ef  :  Symmetry energy
%  26.60.Gj  :  Neutron star crust
%  21.60.-n  :   Nuclear structure models and methods
%  97.10.Sj  :   Pulsations, oscillations, and stellar seismology 
%%%%%%%%%%%%%%%%%%%%%%%%%%%%%%%%%%%%%%%%%%%%%%%%%
%]
% For two column
%}
\maketitle

%\baselineskip 24pt

%%%%%%%%%%%%%%%%%%%%%%%%%%%%%%%%%%%%%%%%%%%%%%%%
\section{Introduction}
\label{sec:I}
%%%%%%%%%%%%%%%%%%%%%%%%%%%%%%%%%%%%%%%%%%%%%%%%

Neutron stars provided via supernova explosion, which is the last moment of massive star's life, are a suitable environment for probing the physics in extreme conditions. The density inside the star easily exceeds the standard nuclear density, while the gravitational and magnetic fields become much stronger than that observed in our solar system. Through the observations of neutron star itself and/or the phenomena associated with the neutron stars, one would obtain the aspect under such extreme conditions. For example, the discovery of the $2M_\odot$ neutron stars \cite{D10,A13,C20} enables us to exclude some of soft equations of state (EOSs), with which the expected maximum mass does not approach the observed neutron star mass. This constraint consequently brings up an issue about the appearance of hyperon inside the neutron stars, the so-called hyperon puzzle. In addition, it is known that the light propagating in the strong gravitational field can bend due to the relativistic effects. By taking into account  this effect, the light curve from a rotating neutron star with hot spot(s) mainly depends on the stellar compactness, which is the ratio of stellar mass to the radius (e.g., \cite{PFC83,LL95,PG03,PO14,SM18,Sotani20a}). In practice, via the X-ray observation by the Neutron star Interior Composition Explorer (NICER) mission, the properties of the millisecond pulsar PSR J0030+0451 has observationally been estimated \cite{Riley19,Miller19}. The further constraints on the neutron star mass and radius will enable us to constrain the EOS for neutron star matter.

Furthermore, the oscillation frequencies of the central objects are also important information. Since the frequency of the objects strongly depends on their interior properties, one would extract such an information via the observation of the corresponding frequency as an inverse problem. This technique is known as seismology on the Earth, helioseismology on the Sun, and asteroseisomogy on astronomical objects. In fact, by identifying the frequencies of the quasi-periodic oscillations observed in the giant flares with the neutron star crustal oscillations, the crust EOS (especially the nuclear saturation parameters) has been constrained (e.g., \cite{GNHL2011,SNIO2012,SIO2016}). In a similar way, once the gravitational waves from compact objects would be observed in the future, one could extract the information about the mass ($M$), radius ($R$), and EOSs for the source objects (e.g., \cite{AK1996,AK1998,STM2001,SH2003,SYMT2011,PA2012,DGKK2013,Sotani20b}). Additionally, in order to identify the origin of gravitational wave signals found in the numerical simulations for core-collapse supernova, the gravitational wave asteroseismology on protoneutron stars newly born via supernova explosions recently attracts attention (e.g., \cite{FMP2003,FKAO2015,ST2016,SKTK2017,MRBV2018,TCPOF19,SS2019}).

Among these attempts, the asteroseismology on cold neutron stars has extensively been studied up to now. For considering the asteroseismology, one first prepares the background models and then makes a linear analysis on the provided background models. Depending on the input physics, the corresponding oscillations would be excited as eigenmodes. That is, if one would identify a frequency from the neutron stars with a specific eigenmode, one could extract the physics behind the oscillation. In particular, the fundamental ($f$-) and the 1st pressure ($p_1$-) modes are observationally important gravitational wave signals from neutron stars composed of perfect fluid, because those frequencies are relatively low among the various eigenmodes \cite{KS1999}. Since the $p$-modes are overtone of the $f$-mode, these modes could be excited in the astrophysical scenarios involving dynamical processes such as neutron star mergers or core-collapse supernovae. A starquake associated with a pulsar glitch may also be considered as another excitation scenario. In addition, if a kind of mini-collapse due to the phase transition inside the neutron star would happen, the $f$- and $p$-modes may be excited. Anyway, since the amplitude of each mode excited in these scenarios is still unknown, which cannot be determined with the linear analysis, one has to carefully estimate it by numerical simulations to discuss the detectability for each gravitational wave  mode.

Meanwhile, it has been also shown that the $f$-mode frequency is strongly associated with the neutron star average density, $M/R^3$. This is because that the $f$-mode is a kind of the acoustic oscillations, whose propagation speed, i.e., sound velocity, is related to the stellar average density. In fact, the $f$-mode frequency is expressed as a linear function of the square root of the stellar average density ($x$) almost independently of the EOS \cite{AK1996,AK1998}, even though there is still a little dependence on the EOS. Moreover, in our previous study for low-mass neutron stars, whose central density is less than two times the nuclear saturation density, we derived the empirical formula for the $f$-mode frequency as a function of $x$ and the parameter $\eta$, which is a combination of the nuclear saturation parameters, mainly adopting the phenomenological EOSs \cite{Sotani20b}. That is, we could clearly separate the EOS dependence from the $x$ dependence in the $f$-mode frequency. Then, we proposed that the value of $\eta$ could be estimated via the observation of the $f$-mode gravitational wave from a low-mass neutron star, whose mass is known, by using the empirical formula for the $f$-mode frequency together with the mass formula derived in Ref.~\cite{SIOO14}. This attempt partially works well, while we also found that it does not work for lower value of $\eta$.

In this study, by adopting the unified realistic EOSs, first we will update the empirical formula for the $f$-mode frequency. With the updated formula, we can overcome the defect in the previous study for estimating lower value of $\eta$. In practice, one could estimate the value of $\eta$ within $\sim 20\%$ accuracy, once the $f$-mode frequency would be observed from the neutron star whose mass is known, if the mass is less than $M=1.174M_\odot$. In addition, we will newly derive the empirical formula for the $p_1$-mode frequency from low-mass cold neutron stars. Then, we will propose that the value of $\eta$ could be estimated via the simultaneous observation of the $f$- and $p_1$-mode gravitational waves by using our empirical formulae. Since the empirical formula for the $p_1$-mode frequencies is only applicable for low-mass neutron stars with the mass less than $\sim 0.9M_\odot$, which is significantly lower than the observed minimum mass of neutrons star, i.e., $M=1.174M_\odot$ for PSR J0453+1559 \cite{Martinez15}, our method is workable only if the neutron stars whose masses are less than $\sim 0.9M_\odot$ would exist. Furthermore, we find the strong correlation between the maximum frequency of the $f$-mode and the minimum radius of neutron star, which corresponds to the neutrons star model with the maximum mass.

This paper is organized as follows. In Sec. \ref{sec:EOS}, we mention the unified EOS considered in this study. In Sec. \ref{sec:Oscillation}, we calculate the eigenfrequencies from the neutron stars constructed with various EOSs. With the resultant eigenfrequencies, we update the empirical formula for the $f$-mode frequency derived in the previous study, and newly derive the empirical formula for the $p_1$-mode frequency. Then, we will show how well the empirical formulae work for estimating the value of $\eta$. Finally, we conclude this study in Sec. \ref{sec:Conclusion}. The details for the update of the empirical formula for the $f$-mode is shown in Appendix \ref{sec:appendix_1}. Unless otherwise mentioned, we adopt geometric units in the following, $c=G=1$, where $c$ denotes the speed of light, and the metric signature is $(-,+,+,+)$.

%%%%%%%%%%%%%%%%%%%%%%%%%%%%%%%%%%%%%%%%%%%%%%%%
\section{Unified EOS considered in this study}
\label{sec:EOS}
%%%%%%%%%%%%%%%%%%%%%%%%%%%%%%%%%%%%%%%%%%%%%%%%

In this study, we simply consider cold neutron stars without rotation as in Ref. \cite{Sotani20b}. The neutron star models are constructed by integrating the Tolman-Oppenheimer-Volkoff equations together with appropriate EOSs. In particular, we adopt the unified realistic EOSs in this study. That is, we consider only  the EOSs, which are consistently constructed with the same framework for both the core and crust region inside the neutron stars. Table \ref{tab:EOS} lists the EOSs considered in this study, where the EOS parameters, such as the incompressibility, $K_0$, the slope parameter for the nuclear symmetry energy, $L$, and an auxiliary parameter, $\eta$, defined as $\eta\equiv (K_0 L^2)^{1/3}$ \cite{SIOO14}, and the maximum mass of the neutron star constructed with each EOS are shown. Here, DD2~\cite{DD2,note_DD2}, Miyatsu~\cite{Miyatsu}, and Shen~\cite{Shen} are the EOSs based on the relativistic framework, FPS~\cite{FPS}, SKa~\cite{SKa}, SLy4~\cite{SLy4}, and SLy9~\cite{SLy9} are the EOSs based on the Skyrme-type effective interaction, and Togashi~\cite{Togashi17} is the EOS derived with variational method.  Owing to the nature of nuclear saturation properties, $K_0$ and $L$ have been more or less constrained via the terrestrial experiments, i.e., $K_0=230\pm 40$ MeV \cite{KM13} and $L=58.9\pm 16$ MeV \cite{Li19}, which leads to the fiducial value of $\eta$ as $70.5 \lsim \eta\lsim 114.8$ MeV.

%\hs{Togashi-kun, can you put here more details about each EOS, as a new paragraph?}

%%%%%%%%%%%%%%%%%%%%%%%%%%%%%%
%   TABLE 1
%%%%%%%%%%%%%%%%%%%%%%%%%%%%%%
\begin{table}
\caption{EOS parameters adopted in this study, the maximum mass, $M_{\rm max}/M_\odot$, for the neutron star constructed with each EOS, and the square root of the normalized stellar average density at the avoided crossing between the $f$- and $p_1$-modes, $x_{\rm AC}$.} 
\label{tab:EOS}
\begin {center}
\begin{tabular}{ccccccc}
\hline\hline
EOS & $K_0$ (MeV) & $L$ (MeV) & $\eta$ (MeV) & $M_{\rm max}/M_\odot$ & $x_{\rm AC}$  \\
\hline
%OI-EOSs
% &   180 & 31.0   & 55.8  & 0.09068  &  \\
% &   180 & 52.2   & 78.9  & 0.07899  &  \\
% &   230 & 42.6   & 74.7  & 0.08637  &  \\
% &   230 & 73.4   & 107   & 0.07345  &  \\
% &   280 & 54.9   & 94.5  & 0.08331  &  \\
% &   280 & 97.5   & 139   & 0.06887  &  \\  % transition densities from cylindrical nuclei to uniform matter
% &   360 & 76.4   & 128   & 0.07918  &  \\ % transition densities from cylindrical-hole nuclei to uniform matter
% &   360 & 146    & 197   & 0.06680  &  \\ % transition densities from spherical nuclei to uniform matter
%\hline
DD2
 & 243 & 55.0  & 90.2  & 2.41 & 0.2085 & \\
Miyatsu
 & 274 &  77.1 & 118  & 1.95 & 0.2097 & \\
Shen
% & 281 & 114  & 154  & \\
 & 281 & 111  & 151  &  2.17  & 0.1886 & \\ \\
%\hline
FPS
 & 261 & 34.9 & 68.2  & 1.80  & 0.2675 & \\ 
SKa
 & 263 & 74.6 & 114 & 2.22 & 0.2021 & \\
SLy4
 & 230 & 45.9 &  78.5 & 2.05 & 0.2345 & \\
SLy9
 & 230 & 54.9 &  88.4 & 2.16 & 0.2263 & \\ \\
%BSk19
% & 237 & 31.9 & 62.3  & \\
%BSk20
% & 241 & 37.4 & 69.6  & \\
%BSk21
% & 246 & 46.6 & 81.1  & \\
%\hline
%APR
% & 266  & 57.6  & 95.9 & 2.17  &  \\
Togashi
 & 245  & 38.7  & 71.6 & 2.21  & 0.2537 &  \\
\hline \hline
\end{tabular}
\end {center}
\end{table}
%%%%%%%%%%%%%%%%%%%%%%%%%%%%%%

%%%%%%%%%%%%%%%%%%%%%%%%%%%%%%%%%%%
% Figure 1
%%%%%%%%%%%%%%%%%%%%%%%%%%%%%%%%%%%
\begin{figure*}[tbp]
\begin{center}
\includegraphics[scale=0.5]{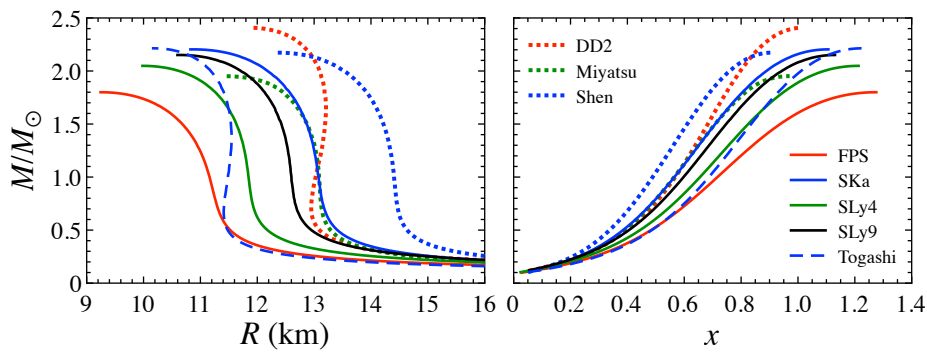}  
\end{center}
\caption{%%
Mass-radius relation for neutron star models constructed with various EOSs is shown in the left panel, where the dotted, solid, and dashed lines correspond to the EOSs based on the relativistic framework, the Skyrme-type effective interactions, and derived with variational method, respectively. In the right panel, the stellar mass is shown as a function of the square root of the normalized stellar average density, $x\equiv (M/1.4M_\odot)^{1/2}(R/10\ {\rm km})^{-3/2}$.
}%%
\label{fig:MR}
\end{figure*}
%%%%%%%%%%%%%%%%%%%%%%%%%%%%%%%%%%%

In the left panel of Fig. \ref{fig:MR}, we show the relation between the mass ($M$) and radius ($R$) of the neutron stars constructed with the EOSs considered in this study, where the dotted, solid, and dashed lines correspond to the EOSs based on the relativistic framework, the Skyrme-type effective interaction, and derived with variational method, respectively. In the right pane of Fig. \ref{fig:MR}, the neutron star mass is shown as a function of the square root of the normalized stellar average density defined as
\begin{equation}
 x\equiv \left(\frac{M}{1.4M_\odot}\right)^{1/2}\left(\frac{R}{10\ {\rm km}}\right)^{-3/2}. 
\end{equation} 
In the both panels, we plot the neutron star models up to that with the maximum mass. From this figure, it is clearly observed  that the neutron star model with the maximum mass has the minimum radius and the maximum value of $x$ for each EOS.

The advantage for considering the unified EOSs is that the properties of low-mass neutron stars can be characterized well as a function of $\eta$ \cite{SIOO14}. In fact, the mass and gravitational redshift for low-mass neutron stars are written as a function of $\eta$ and the central density normalized by the saturation density, $\rho_0$. In addition, it has been shown that the properties for slowly rotating low-mass neutron stars are associated with $\eta$ \cite{SSB16} and that the possible maximum mass of neutron star is also discussed with $\eta$ \cite{Sotani17,SK17}. Furthermore, we have shown that $x$ for low-mass neutron stars is expressed as a function of $\eta$ \cite{Sotani20b}, which is updated in Appendix \ref{sec:appendix_1} by using the EOSs considered in this study, i.e., Eqs. (\ref{eq:x-eta}) - (\ref{eq:c2}).

%%%%%%%%%%%%%%%%%%%%%%%%%%%%%%%%%%%%%%%%%%%%%%%%
\section{Gravitational wave asteroseismology}
\label{sec:Oscillation}
%%%%%%%%%%%%%%%%%%%%%%%%%%%%%%%%%%%%%%%%%%%%%%%%

On the neutron star models constructed with the various EOSs, the gravitational wave frequencies are determined via linear analysis. In this study we simply adopt the relativistic Cowling approximation, where the metric perturbations are neglected during the fluid oscillations. The perturbation equations are derived by linearizing the energy-momentum conservation law. Then, imposing the appropriate boundary conditions at the stellar center and surface, the problem to solve becomes an eigenvalue problem, where the eigenvalue, $\omega$, is associated with the gravitational wave frequency, $f$, via $\omega=2\pi f$. The concrete perturbation equations and boundary conditions are the same as in Ref. \cite{SYMT2011}.
In this study, we focus on only the $\ell=2$ modes, because the $\ell=2$ modes are considered to be more energetic in gravitational wave radiation. We remark that one can qualitatively discuss the behavior of the eigenfrequencies even with the relativistic Cowling approximation, while the $f$- and $p$-modes calculated with the approximation deviate from the frequencies calculated with metric perturbations (without relativistic Cowling approximation)  within $\sim 20\%$ and $\sim 10\%$, respectively \cite{YK1997}.

%%%%%%%%%%%%%%%%%%%%%%%%%%%%%%%%%%%
% Figure 2
%%%%%%%%%%%%%%%%%%%%%%%%%%%%%%%%%%%
\begin{figure}[tbp]
\begin{center}
\includegraphics[scale=0.5]{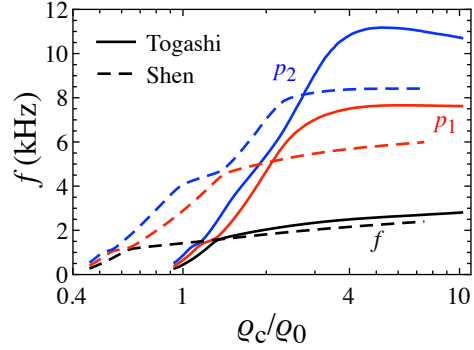}  
\end{center}
\caption{%%
The $f$-, $p_1$-, and $p_2$-mode frequencies are shown as a function of the central density, $\rho_c$, normalized by the saturation density, $\rho_0$, for the neutron star models constructed with Togashi EOS (solid lines) and Shen EOS (dashed lines).
}%%
\label{fig:f-uc-TS}
\end{figure}
%%%%%%%%%%%%%%%%%%%%%%%%%%%%%%%%%%%

%%%%%%%%%%%%%%%%%%%%%%%%%%%%%%%%%%%
% Figure 3
%%%%%%%%%%%%%%%%%%%%%%%%%%%%%%%%%%%
\begin{figure}[tbp]
\begin{center}
\includegraphics[scale=0.5]{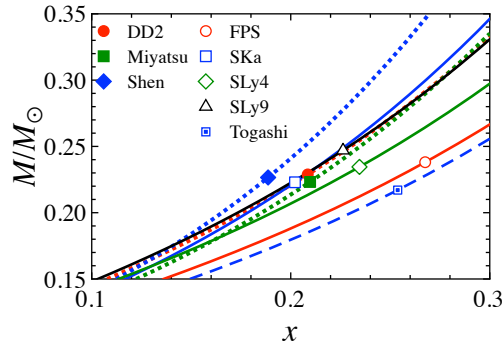}  
\end{center}
\caption{%%
In the relation between $x$ and $M/M_\odot$, which is just extended figure for the right panel of Fig. \ref{fig:MR}, the stellar models for the avoided crossing between the $f$- and $p_1$-modes for various EOSs are shown with marks. 
}%%
\label{fig:Mx1}
\end{figure}
%%%%%%%%%%%%%%%%%%%%%%%%%%%%%%%%%%%

Since we consider the gravitational waves from cold neutron stars composed of the perfect fluid in this study, only the $f$- and $p_i$-modes can be excited. In Fig. \ref{fig:f-uc-TS}, as an example we show the $f$-, $p_1$-, and $p_2$-modes as a function of the central density $\rho_c$ normalized by the nuclear saturation density $\rho_0$ for the neutron star models constructed with Togashi EOS (solid lines) and Shen EOS (dashed lines). From this figure, one can observe that the central density when the avoided crossing happens strongly depends on the EOS. Even so, we have shown in the previous study \cite{Sotani20b} that the $f$-mode frequency, $f_{f,{\rm AC}}$, and the square root of the normalized stellar average density, $x_{\rm AC}$, for the neutron star model at the avoided crossing between the $f$- and the $p_1$-modes are expressed well as a function of $\eta$, which are updated as Eqs. (\ref{eq:ffAC}) and (\ref{eq:xAC}) by using the EOSs considered in this study in Appendix \ref{sec:appendix_1}. 
That is, $x_{\rm AC}$ depends on the EOS. The values of $x_{\rm AC}$ for various EOSs are shown in Table \ref{tab:EOS}, which are also shown in Fig. \ref{fig:Mx1} with marks. 
Additionally, using the dependence of $f_{f,{\rm AC}}$ and $x_{\rm AC}$ on $\eta$, we have also derived the empirical formula for the $f$-mode frequencies for low-mass neutrons stars as a function of $x$ and $\eta$. This empirical formula is also updated with the EOSs considered in this study (see in Appendix \ref{sec:appendix_1} for the details), such as
\begin{equation}
  f_f(x,\eta)\ {\rm (kHz)}  = 1.6970x + f_0(\eta), \label{eq:ff_x}
\end{equation}
where $f_0(\eta)$ is given by Eq. (\ref{eq: ff0}).

\begin{figure}[tbp]
\begin{center}
\includegraphics[scale=0.5]{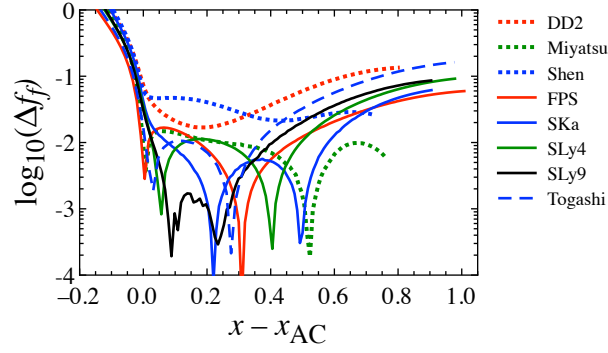}  
\end{center}
\caption{%%
Relative deviation, $\Delta f_f$, of the $f$-mode frequency estimated with Eq. (\ref{eq:ff_x}) from the $f$-mode frequency calculated for each neutron star model is shown as a function of $x-x_{\rm AC}$ for various EOSs. $\Delta f_f$ is calculated with Eq. (\ref{eq:abs_dff}).
}%%
\label{fig:abs_df}
\end{figure}

In the previous study \cite{Sotani20b}, we have proposed that the value of $\eta$ could be estimated by the observation of the $f$-mode frequency from the neutron star whose mass is known, using the empirical formula for the $f$-mode frequency together with the mass formula derived in Ref. \cite{SIOO14}, where the estimation of lower value of $\eta$ does not work well with the original empirical formula for the $f$-mode. Owing to the update in this study, we find that this defect can be removed (see Fig. \ref{fig:Delta} and in Appendix \ref{sec:appendix_1} for the details). In addition, we considered only the low-mass neutron stars whose central density is less than $2\rho_0$ in the previous study, but we find that the empirical formula given by Eq. (\ref{eq:ff_x}) is applicable even for canonical neutron stars. To check the accuracy of this formula, we calculate the relative deviation of the $f$-mode frequency estimated with Eq. (\ref{eq:ff_x}) from the frequency calculated for each stellar model, $f_f$, via
\begin{equation}
  \Delta f_f = \frac{|f_f - f_f(x,\eta)|}{f_f}, \label{eq:abs_dff}
\end{equation}
and the result is shown in Fig. \ref{fig:abs_df}. From this figure, we find that the $f$-mode frequency even for the neutron stars with maximum mass can be estimated with Eq. (\ref{eq:ff_x}) less than $20\%$ accuracy.

%%%%%%%%%%%%%%%%%%%%%%%%%%%%%%%%%%%
% Figure 4
%%%%%%%%%%%%%%%%%%%%%%%%%%%%%%%%%%%
%%%%%%%%%%%%%%%%%%%%%%%%%%%%%%%%%%%

\begin{figure*}[tbp]
\begin{center}
\includegraphics[scale=0.45]{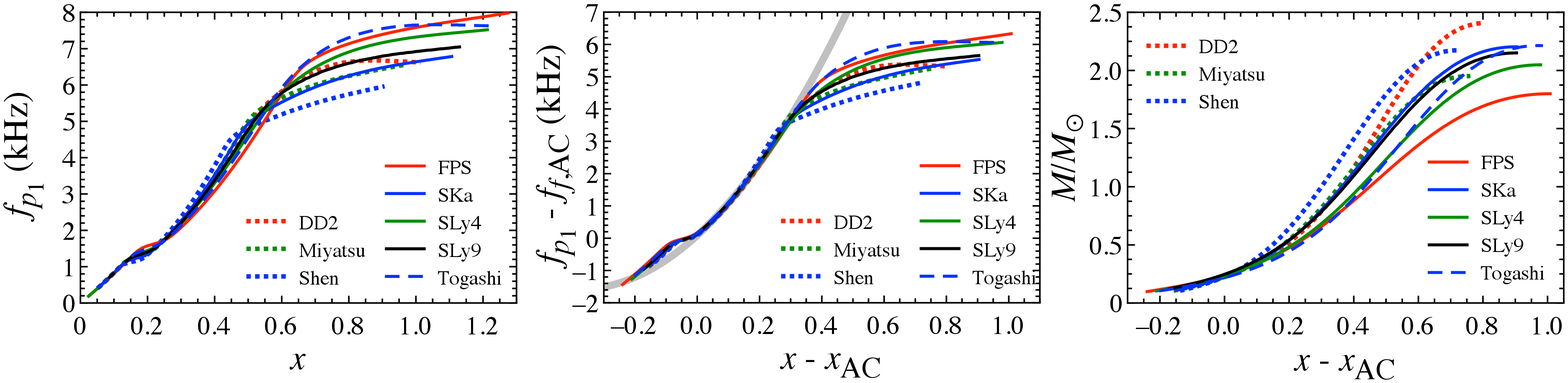}   
\end{center}
\caption{%%
The $p_1$-mode frequencies from the neutron star models constructed with various EOSs are shown as a function of $x$ in the left panel. In the middle panel, the shifted $p_1$-mode frequencies, $f_{p_1}-f_{f,{\rm AC}}$, are shown as a function of the shifted square root of the normalized stellar average density, $x-x_{\rm AC}$, where the thick-solid line denotes the fitting line given by Eq. (\ref{eq:fp1_xx}). In the right panel, the neutron star mass is shown as a function of $x-x_{\rm AC}$.
}%%
\label{fig:fp1-xx}
\end{figure*}

On the other hand, in the left panel of Fig. \ref{fig:fp1-xx} we show the $p_1$-mode frequency as a function of $x$. One can observe three avoided crossing in the $p_1$-mode, i.e., the first with the $p_2$-mode, the second with the $f$-mode, and the third with the $p_2$-mode again (see also Fig. \ref{fig:f-uc-TS}), as the central density increases. Since the $p_1$-mode frequency and the value of $x$ for the neutron star model at the second avoided crossing (with the $f$-mode) are strongly associated with $f_{f,{\rm AC}}$ and $x_{\rm AC}$, in the middle panel of Fig. \ref{fig:fp1-xx} we show the $p_1$-mode frequency shifted by $f_{f,{\rm AC}}$ as a function of $x-x_{\rm AC}$. From this figure, $f_{p_1}-f_{f,\rm AC}$ in the region between the second and third avoided crossing is written as a function of $x-x_{\rm AC}$ almost independently of the EOS, where the EOS dependence in this region is confined to $f_{f,{\rm AC}}$ and $x_{\rm AC}$. In practice, we derive the fitting line in this region as 
\begin{equation}
  f_{p_1} - f_{f,{\rm AC}}\ {\rm (kHz)}  = 11.5877(x-x_{\rm AC})^2 + 8.8014(x-x_{\rm AC}) + 0.087307, 
     \label{eq:fp1_xx}
\end{equation}
which is also shown with the thick-solid line in the middle panel of Fig. \ref{fig:fp1-xx}. Since $f_{f,{\rm AC}}$ and $x_{\rm AC}$ in this equation are expressed as a function of $\eta$, i.e., Eqs. (\ref{eq:ffAC}) and (\ref{eq:xAC}), we eventually obtain $f_{p_1}$ as a function of $x$ and $\eta$ (although the explicit function form is not shown here), such as
\begin{equation}
  f_{p_1}=f_{p_1}(x,\eta),  \label{eq:fp1_xx1}
\end{equation}
even though this fitting formula can be valid only for rather light neutron stars whose masses are less than $\sim 0.9M_\odot$, as seen the right panel of Fig. \ref{fig:fp1-xx}.

\begin{figure}[tbp]
\begin{center}
\includegraphics[scale=0.5]{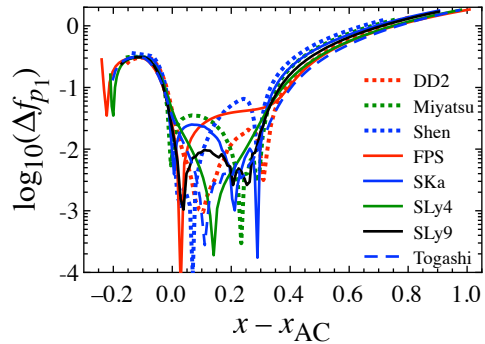}  
\end{center}
\caption{%%
Relative deviation, $\Delta f_{p_1}$, of the $p_1$-mode frequency estimated with Eq. (\ref{eq:fp1_xx1}) from the $p_1$-mode frequency calculated for each neutron star model is shown as a function of $x-x_{\rm AC}$ for various EOSs. $\Delta f_{p_1}$ is calculated with Eq. (\ref{eq:abs_dfp1}).
}%%
\label{fig:abs_dfp1}
\end{figure}

In order to check the accuracy of the obtained empirical formula for the $p_1$-mode frequency, in a similar way to the case of the $f$-mode, we calculate the relative deviation via
\begin{equation}
  \Delta f_{p_1} = \frac{|f_{p_1} - f_{p_1}(x,\eta)|}{f_{p_1}}, \label{eq:abs_dfp1}
\end{equation}
where $f_{p_1}$ denotes the $p_1$-mode frequency calculated for each stellar model, while $f_{p_1}(x,\eta)$ denotes the frequency estimated with the empirical formula given by Eq. (\ref{eq:fp1_xx1}) by using the value of $x$ and $\eta$ for each stellar model. The resultant value of $\Delta f_{p_1}$ is shown in Fig. \ref{fig:abs_dfp1} as a function of $x-x_{\rm AC}$. Since the $p_1$-mode frequencies for more massive  neutron stars than that at the third avoided crossing with the $p_2$-mode significantly deviate from the empirical formula given by Eq. (\ref{eq:fp1_xx1}) as shown in the right panel of Fig. \ref{fig:fp1-xx}, the accuracy becomes worse for the massive neutron star models. Meanwhile, we find that the $p_1$-mode frequencies for the neutron star models between the second and third avoided crossing are well estimated with our empirical formula.

%%%%%%%%%%%%%%%%%%%%%%%%%%%%%%%%%%%
% Figure 5
%%%%%%%%%%%%%%%%%%%%%%%%%%%%%%%%%%%
%%%%%%%%%%%%%%%%%%%%%%%%%%%%%%%%%%%

%%%%%%%%%%%%%%%%%%%%%%%%%%%%%%%%%%%
% Figure 6
%%%%%%%%%%%%%%%%%%%%%%%%%%%%%%%%%%%
%%%%%%%%%%%%%%%%%%%%%%%%%%%%%%%%%%%

\begin{figure}[tbp]
\begin{center}
\includegraphics[scale=0.5]{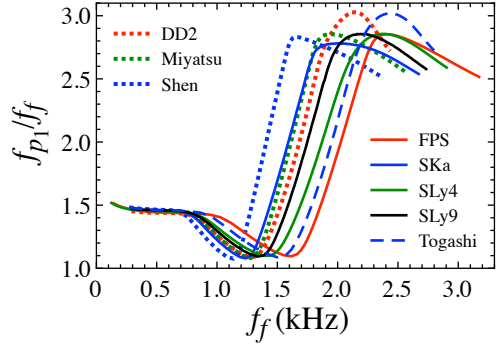}  
\end{center}
\caption{%%
For the neutron star models constructed with various EOSs, the ratio of the $p_1$-mode frequency to the $f$-mode frequency, $f_{p_1}/f_f$, is shown as a function of the $f$-mode frequency. The minimum of $f_{p_1}/f_f$ corresponds to the neutron star model, where the avoided crossing arises between the $f$- and the $p_1$-modes (the second avoided crossing), while the maximum of $f_{p_1}/f_f$ is to the neutron star model, where the avoided crossing arises between the $p_1$- and the $p_2$-modes (the third avoided crossing).
}%%
\label{fig:f-p1f}
\end{figure}

\begin{figure*}[tbp]
\begin{center}
\begin{tabular}{cc}
\includegraphics[scale=0.5]{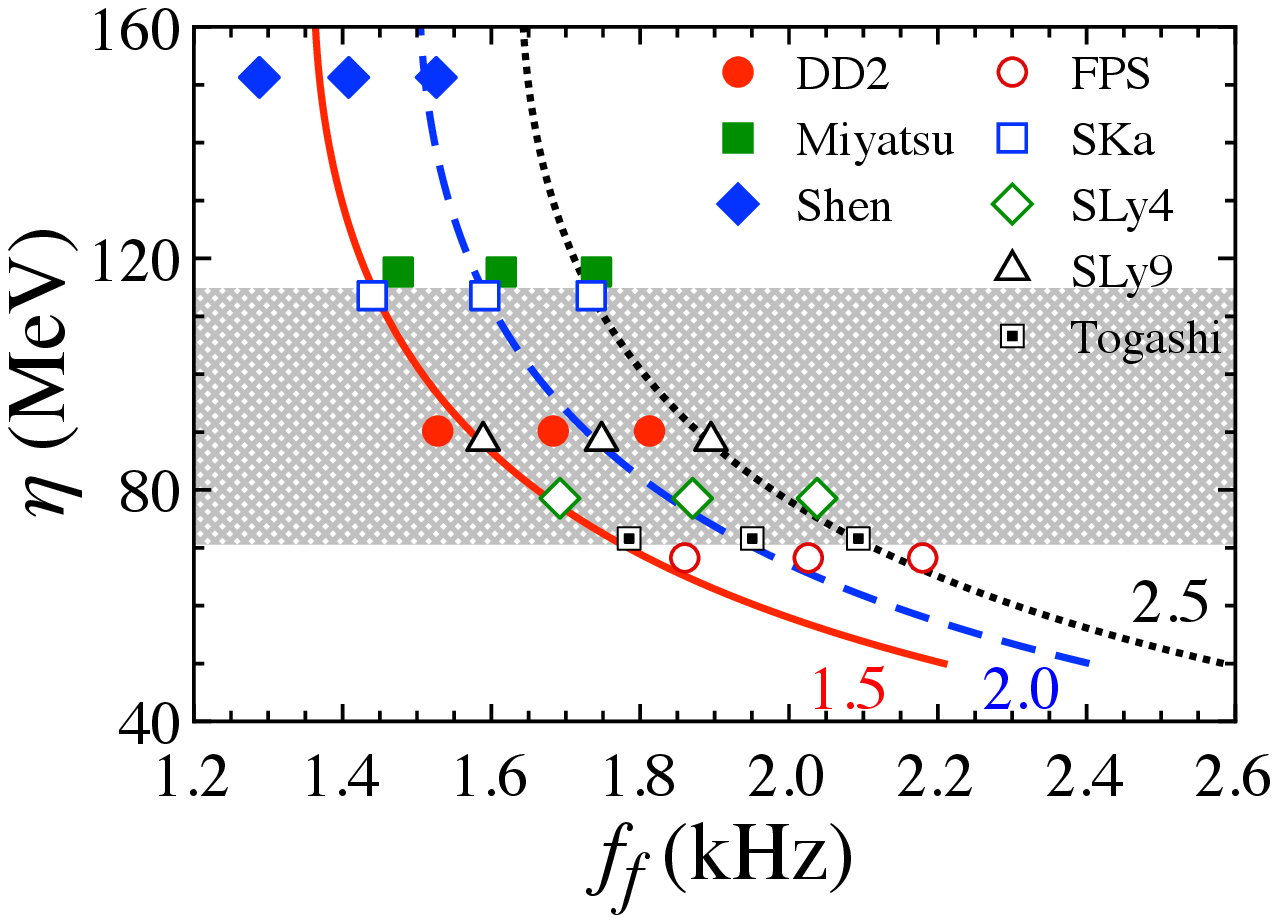} &
\includegraphics[scale=0.5]{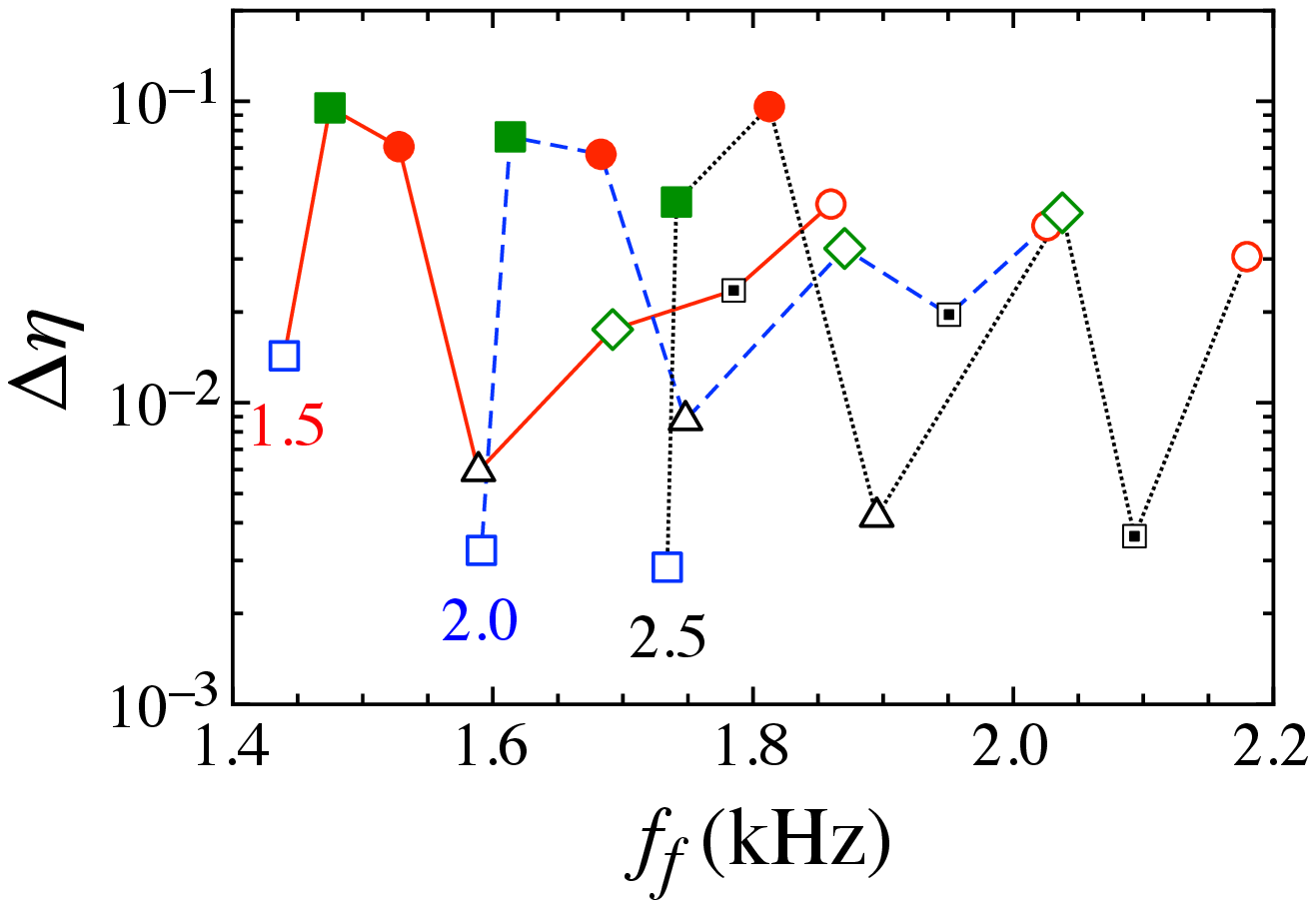}  
\end{tabular}
\end{center}
\caption{%%
In the left panel, by fixing the value of $f_{p_1}/f_f$, where the $f$- and $p_1$-mode frequencies, $f_f$ and $f_{p_1}$, are estimated with the empirical formulae given by Eqs. (\ref{eq:ff_x}) and (\ref{eq:fp1_xx1}), $\eta(f_f,f_{p_1})$ estimated with the empirical formula is shown as a function of $f_f$, where the solid, dashed, and dotted lines correspond to the results for $f_{p_1}/f_f=1.5$, 2.0, and 2.5, respectively. On this figure, the frequencies calculated for each stellar model are also shown with marks. For reference, the fiducial value of $\eta$ constrained by terrestrial experiments is shown by shaded region. In the right panel, the relative deviation of $\eta$ estimated with the empirical formulae from the value of $\eta$ for each stellar model is shown as a function of the $f$-mode frequency, where the meaning of marks is the same as in the left panel, while the solid, dashed, and dotted lines correspond to the case with $f_{p_1}/f_f=1.5$, 2.0, and 2.5. The relative deviation is calculated with Eq. (\ref{eq:Delta_eta}).
}%%
\label{fig:Delta-eta}
\end{figure*}

Now, we have derived two empirical formulae for the $f$- and $p_1$-mode frequencies as a function of $x$ and $\eta$, i.e., Eqs. (\ref{eq:ff_x}) and (\ref{eq:fp1_xx1}). With using the empirical formula for the $f$-mode frequency, the $p_1$-mode can be estimated as a function of $f_f$ and $\eta$, i.e., $f_{p_1}=f_{p_1}(f_f,\eta)$, with which one can in principle rewrite as $\eta=\eta(f_f,f_{p_1})$. That is, one could constrain the value of $\eta$ via the simultaneous observation of the $f$- and $p_1$-mode gravitational waves from neutron stars. In fact, for the neutron star models between the second and third avoided crossing in the $p_1$-mode, as shown in Fig. \ref{fig:f-p1f} the ratio of the $p_1$- to the $f$-modes are almost linear function of the $f$-mode frequency and that relation seems to be ordered by the value of $\eta$, i.e., $\eta$ increases from right to left. In order to see how accurate the value of $\eta$ would be estimated by using the empirical formulae via the simultaneous observation of the $f$- and $p_1$-modes, we show the value of $\eta(f_f,f_{p_1})$ estimated with the empirical formula as a function of the $f$-mode frequency for the case with $f_{p_1}/f_f=1.5$ (solid line), 2.0 (dashed line), and 2.5 (dotted line) in the left panel of Fig. \ref{fig:Delta-eta}, while the $f$-mode frequencies calculated for each stellar model (having own value of $\eta$) are also plotted with marks. We find that the value of $\eta$ is well estimated with the empirical formulae for the $f$- and $p_1$-modes except for the case with Shen EOS. Nevertheless, by considering the fact that the nuclear saturation parameters for Shen EOS have been excluded by the terrestrial experiments and that the neutron star models constructed with Shen EOS are also excluded by the gravitational wave observation in the GW170817 event \cite{Annala18}, maybe one should take care of the EOSs except for Shen EOS. So, with respect to the results shown in the left panel of Fig. \ref{fig:Delta-eta} except for Shen EOS, the relative deviation of the value of $\eta$ estimated with the empirical formulae from the value of $\eta$ for each neutron star model is evaluated via
\begin{equation}
  \Delta\eta = \frac{|\eta - \eta(f_f,f_{p_1})|}{\eta}, \label{eq:Delta_eta}
\end{equation}
which is shown as a function of the $f$-mode frequency in the right panel of Fig. \ref{fig:Delta-eta}, where the solid, dashed, and dotted lines correspond to the results for $f_{p_1}/f_f=1.5$, 2.0, and 2.5, respectively. As a result, we find that via the simultaneous observation of the $f$- and $p_1$-mode gravitational waves for $1.5\lsim f_{p_1}/f_f\lsim 2.5$, one could extract the value of $\eta$ less than $\sim 10\%$ accuracy. We remark that, since in Fig. \ref{fig:f-p1f} the minimum and maximum of $f_{p_1}/f_f$ correspond to the neutron star models at the second and third avoided crossing in the $p_1$-mode, maybe one can estimate the value of $\eta$ even for $1.1\lsim f_{p_1}/f_f\lsim 1.5$ and $2.5\lsim f_{p_1}/f_f\lsim 2.7$ with using our empirical formulae. But, in such a case since the $f$-mode frequency has two solutions with the fixed value of $f_{p_1}/f_f$, we only focus on the case with $1.5\lsim f_{p_1}/f_f\lsim 2.5$ in this study.

%%%%%%%%%%%%%%%%%%%%%%%%%%%%%%%%%%%
% Figure 7
%%%%%%%%%%%%%%%%%%%%%%%%%%%%%%%%%%%
%%%%%%%%%%%%%%%%%%%%%%%%%%%%%%%%%%%

%%%%%%%%%%%%%%%%%%%%%%%%%%%%%%%%%%%
% Figure 8
%%%%%%%%%%%%%%%%%%%%%%%%%%%%%%%%%%%
%%%%%%%%%%%%%%%%%%%%%%%%%%%%%%%%%%%

\begin{figure}[tbp]
\begin{center}
\includegraphics[scale=0.5]{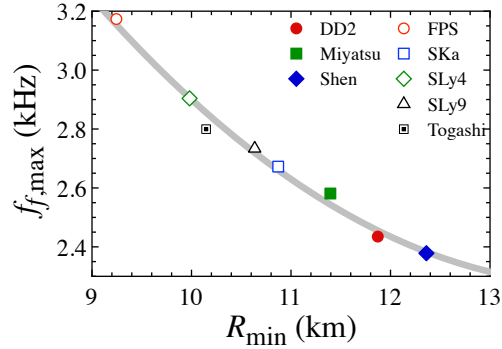}  
\end{center}
\caption{%%
The maximum $f$-mode frequency for each EOS is shown as a function of the minimum radius of neutron star, which corresponds to the neutron star model with the maximum mass. The thick-solid line is fitting line given by Eq. (\ref{eq:ff-Rmin}).
}%%
\label{fig:ff-Rmin}
\end{figure}

Furthermore, as shown in Fig. \ref{fig:ff-Rmin} we also find that the maximum $f$-mode frequency, which comes from the neutron star model with the maximum mass, is strongly correlated to the radius of the neutron star model with the maximum mass, which corresponds to the minimum radius of neutron star, such as
\begin{equation}
  f_{f,{\rm max}}\ {\rm (kHz)} =  3.7739 \left(\frac{R_{\rm min}}{10\ {\rm km}}\right)^2 
       -10.6289\left(\frac{R_{\rm min}}{10\ {\rm km}}\right) + 9.7550. \label{eq:ff-Rmin}
\end{equation}
That is, if one would observe a large frequency of the $f$-mode gravitational wave from a cold neutron star, one can constrain the upper limit of the minimum neutron star radius, which enables us to exclude some of stiff EOSs.

%%%%%%%%%%%%%%%%%%%%%%%%%%%%%%%%%%%
% Figure 9
%%%%%%%%%%%%%%%%%%%%%%%%%%%%%%%%%%%
%%%%%%%%%%%%%%%%%%%%%%%%%%%%%%%%%%%

%%%%%%%%%%%%%%%%%%%%%%%%%%%%%%%%%%%%%%%%%%%%%%%%
\section{Conclusion}
\label{sec:Conclusion}
%%%%%%%%%%%%%%%%%%%%%%%%%%%%%%%%%%%%%%%%%%%%%%%%

We examined the $f$- and $p_1$-mode frequencies from cold neutron stars constructed various unified EOSs. First, by using the realistic EOSs considered in this study, we updated the empirical formula for the $f$-mode frequency as a function of the square root of the normalized stellar average density, $x$, and a parameter, $\eta$, which is a combination of the nuclear saturation parameters. The similar empirical formula has already been derived in the previous study for low-mass neutron stars by mainly using the phenomenological EOS, where we proposed that the value of $\eta$ would observationally be estimated with the empirical formula for the $f$-mode together with the mass formula, but it does not work well in the estimation of lower value of $\eta$. Owing to the update in this study, not only this defect can be removed, but also we confirmed that the updated empirical formula is applicable within $\sim 20\%$ accuracy even for the neutron star model with the maximum mass. In addition, we newly derived the empirical formula for the $p_1$-mode frequency as a function of $x$ and $\eta$. As a result, we proposed that the value of $\eta$ could be estimated with the simultaneous observation of the $f$- and $p_1$-modes by using our empirical formulae. In fact, according to our proposal, we showed that $\eta$ could be estimated within $\sim 10\%$ accuracy, when the ratio of the $p_1$-mode to the $f$-mode is in the region between 1.5 and 2.5. Finally, we also found that the maximum $f$-mode frequency is strongly associated with the minimum radius of the neutron star (which corresponds to the stellar model with the maximum mass), and consequently derived the fitting formula for the maximum $f$-mode frequency as a function of the minimum stellar radius. That is, if one would observe a large frequency of the $f$-mode, one might  constrain the upper limit of the minimum stellar radius, which helps us to understand the EOS for high density region. 
In this study we only consider the non-rotating neutron star models. That is, we do not know whether or not our approach can be applied even for rotating neutron stars. This is a very interesting topics and we will study somewhere in the future.

%\newpage
%%%%%%%%%%%%%%%%%%%%%%%%%%%%%%%%%%%%%%%%%%%%%%%%
\acknowledgments
%%%%%%%%%%%%%%%%%%%%%%%%%%%%%%%%%%%%%%%%%%%%%%%%

We would like to thank H. Togashi for providing some of EOS data and also K. Iida for valuable comments.
This work is supported in part by Japan Society for the Promotion of Science (JSPS) KAKENHI Grant Numbers JP18H05236, JP19KK0354, and JP20H04753 and by Pioneering Program of RIKEN for Evolution of Matter in the Universe (r-EMU).

\appendix
%%%%%%%%%%%%%%%%%%%%%%%%%%%%%%%%%%%%%%%%%%%%%%%%
\section{Empirical formula for the $f$-mode frequency from low-mass neutron stars}   % Appendix A
\label{sec:appendix_1}
%%%%%%%%%%%%%%%%%%%%%%%%%%%%%%%%%%%%%%%%%%%%%%%%

\begin{figure}[tbp]
\begin{center}
\includegraphics[scale=0.5]{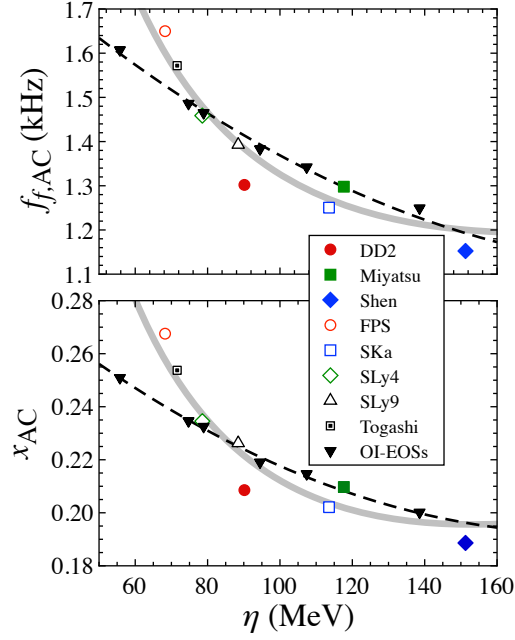}  
\end{center}
\caption{%%
The $f$-mode frequency (top panel) and the square root of the normalized stellar average density (bottom panel) for the neutron star model at the avoided crossing between the $f$- and $p_1$-modes are shown as a function of $\eta$. For reference, we also show the results with OI-EOSs obtained in Ref.~\cite{Sotani20b}. The dashed lines correspond to the fitting formula derived in Ref. \cite{Sotani20b}, which are Eqs. (10) and (14) in Ref.~\cite{Sotani20b}, while the thick-solid lines are the fitting formula updated in this study, i.e., Eqs. (\ref{eq:ffAC}) and (\ref{eq:xAC}).
}%%
\label{fig:fAC-eta}
\end{figure}

\begin{figure*}[tbp]
\begin{center}
\begin{tabular}{cc}
\includegraphics[scale=0.5]{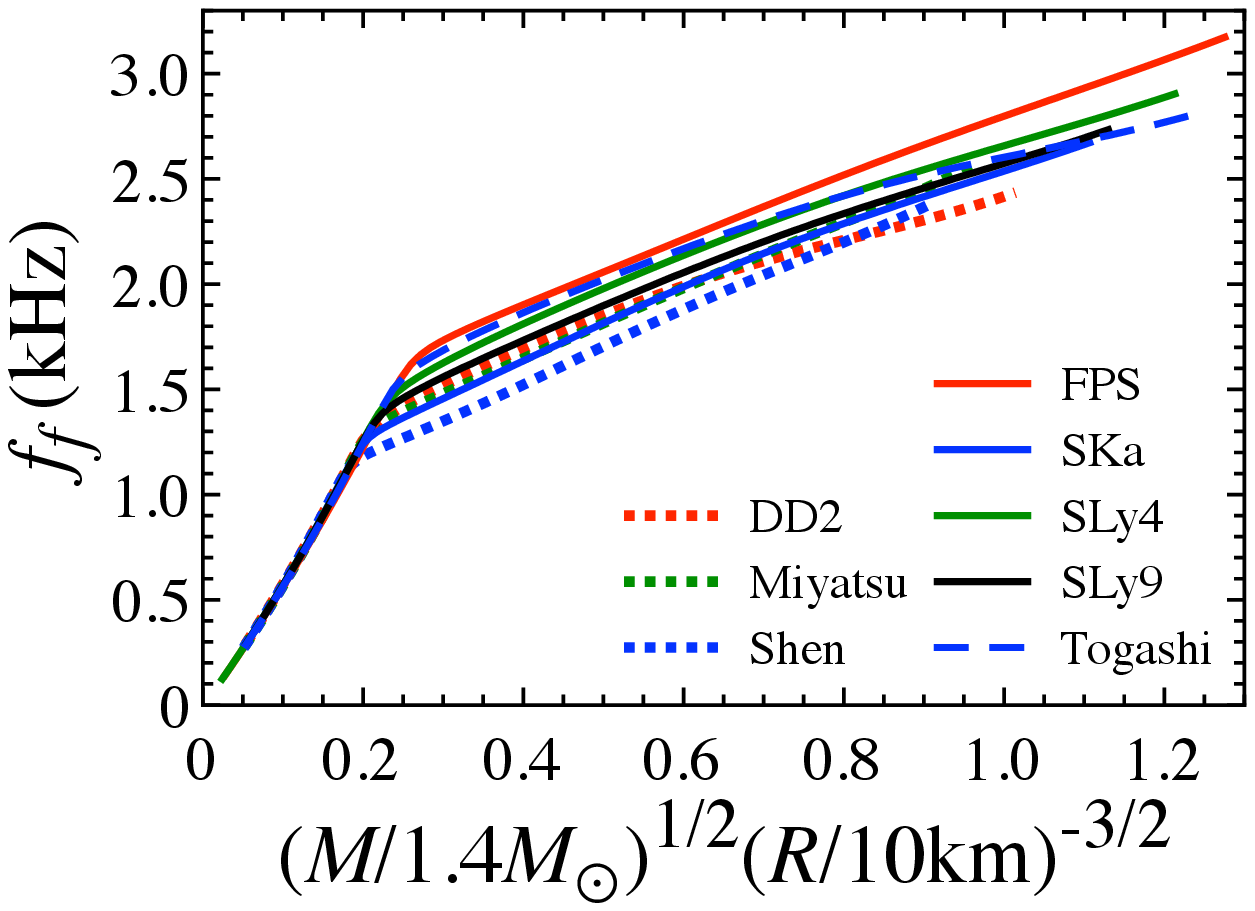} &
\includegraphics[scale=0.5]{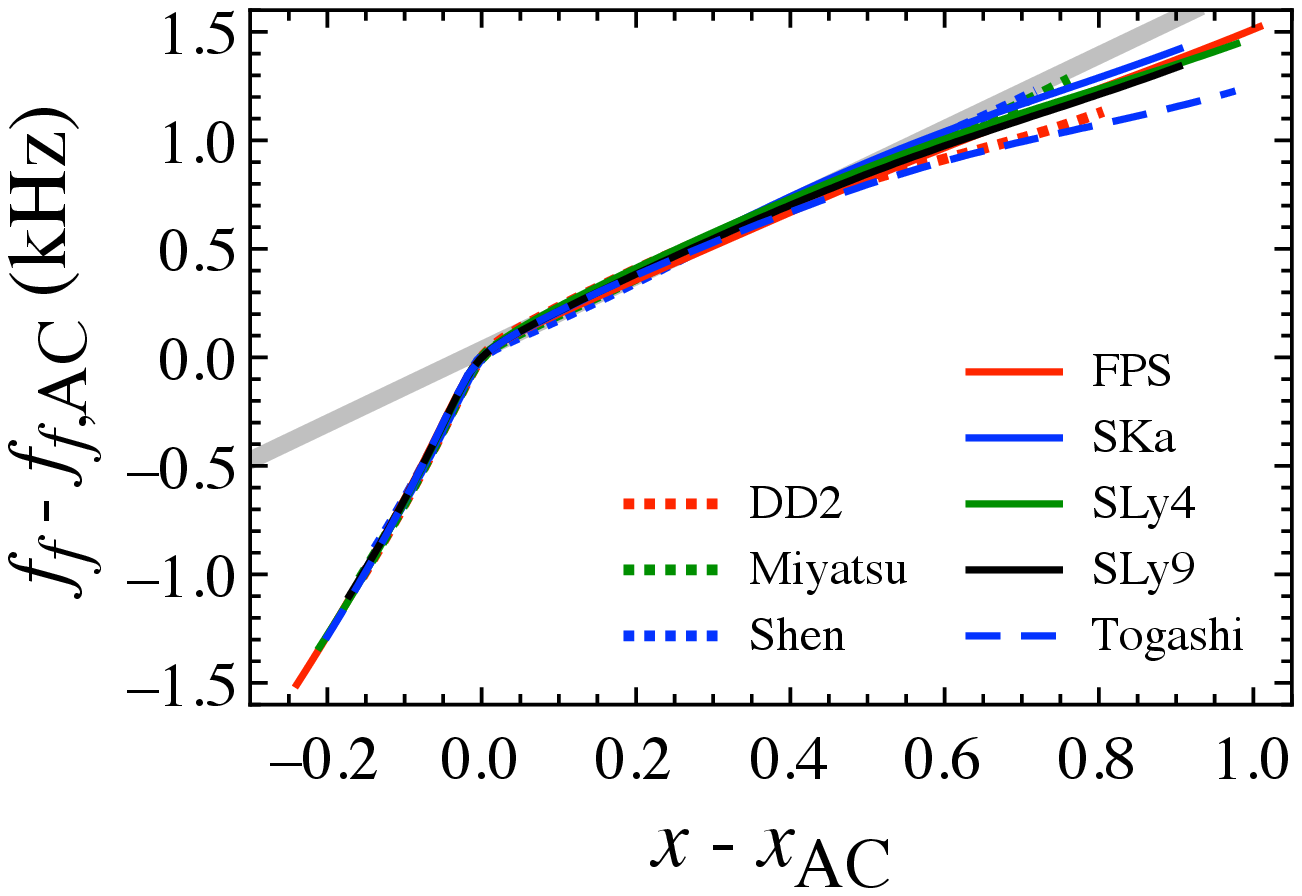}  
\end{tabular}
\end{center}
\caption{%%
The $f$-mode frequencies for various EOSs are shown as a function of the square root of the normalized stellar average density in the left panel. The  shifted $f$-mode frequencies, $f_f-f_{f,{\rm AC}}$, are shown as a function of the shifted square root of the normalized stellar average density, $x-x_{\rm AC}$, in the right panel. In the right panel, the thick-solid line denotes the fitting formula given by Eq. (\ref{eq: ff-xx0}).
}%%
\label{fig:ffAC}
\end{figure*}

With respect to low-mass neutron stars, we have already derived the empirical formula expressing the $f$-mode frequencies in Ref.~\cite{Sotani20b}, where we mainly discussed with the phenomenological EOSs  (the so-called OI-EOSs) proposed in Refs. \cite{OI03,OI07}. In this appendix we try to update it, taking into account the results with the unified realistic EOSs considered in this study. In this procedure, we will omit the case with OI-EOS with $K_0=180$ and $L=31.0$ MeV, which corresponds to $\eta=55.8$ MeV, although we included this parameter set in the previous study~\cite{Sotani20b}, because this parameter set has been excluded by the terrestrial experiments as mentioned in text. As discussed in Refs.~\cite{AK1996,AK1998}, the $f$-mode frequencies are characterized by the neutron star average density almost independently of the EOSs, but the frequencies still depend on the EOS a little (see the left panel of Fig.~\ref{fig:ffAC}). This is because of the phenomena of the avoided crossing between the $f$- and the $p_1$-mode frequencies, where the corresponding stellar model and the frequencies depend on the EOSs. Even so, we have found that the $f$-mode frequencies, $f_{f,{\rm AC}}$, and the square root of  the normalized stellar average density, $x_{\rm AC}$, at the avoided crossing can be expressed well as a function of $\eta$, which is a nuclear parameter characterizing the EOSs. In fact, as in Fig.~\ref{fig:fAC-eta}, the fitting formula for $f_{f,{\rm AC}}$ and $x_{\rm AC}$ are updated as 
\begin{gather}
  f_{f,{\rm AC}}\ ({\rm kHz}) = 0.7705 \eta_{100}^{-1} +0.2912 +0.2642\eta_{100} , \label{eq:ffAC} \\
  x_{\rm AC} =  0.1376 \eta_{100}^{-1} +0.01600  +0.05860\eta_{100},  \label{eq:xAC}
\end{gather}
where $\eta_{100}\equiv \eta/100\ {\rm MeV}$. We remark that, in order to find the central density, $\rho_{c, {\rm AC}}$, for the stellar model at the avoided crossing, the $f$-mode frequencies in the both region with lower and higher density than $\rho_{c, {\rm AC}}$ are fitted with the function form given by
\begin{equation}
  f_f\ {\rm (kHz)} = a_1 + a_2(\log u_c) + a_3(\log u_c)^2 + a_4(\log u_c)^3 \label{eq:AC0}
\end{equation}
in this study, instead of the function form adopted in Ref.~\cite{Sotani20b} as
\begin{equation}
  f_f\ {\rm (kHz)} = a_1 + a_2u_c + a_3u_c^2 + a_4u_c^3,    \label{eq:AC1}
\end{equation}
where $u_c$ is the central density, $\rho_c$, normalized by the saturation density, $\rho_0$, i.e., $u_c\equiv \rho_c/\rho_0$. Then, $\rho_{c, {\rm AC}}$ is identified as an intersection of the fitting in the both region. Anyway, the central density for the stellar model at the avoided crossing almost independent of the function form given by either Eq. (\ref{eq:AC0}) or (\ref{eq:AC1}). 

%%%%%%%%%%%%%%%%%%%%%%%%%%%%%%%%%%%
% Figure A1
%%%%%%%%%%%%%%%%%%%%%%%%%%%%%%%%%%%
%%%%%%%%%%%%%%%%%%%%%%%%%%%%%%%%%%%

With using the realistic EOSs considered in this study, the $f$-mode frequencies are shown as a function of the square root of the normalized stellar average density in the left panel of Fig.~\ref{fig:ffAC}. As mentioned before, the $f$-mode frequencies depend on the EOSs a little, where the difference becomes $\sim 0.5$ kHz. In a similar way to Ref.~\cite{Sotani20b}, the $f$-mode frequencies and the square root of the normalized stellar average density are shifted in such a way that the corresponding values at the avoided crossing are aligned at the origin as in the right panel of Fig.~\ref{fig:ffAC}. The thick-solid line in this figure is the fitting line for $f_f-f_{f,{\rm AC}}$ as a linear function of $x-x_{\rm AC}$, using the data in the region from $x-x_{\rm AC}=0$ up to $\sim 0.4$, which is given by 
\begin{equation}
    f_f - f_{f,{\rm AC}}\ {\rm (kHz)} = 1.6970(x-x_{\rm AC}) + 0.034816. \label{eq: ff-xx0}
\end{equation}
Then, substituting the fitting formulae for $f_{f,{\rm AC}}$ and $x_{\rm AC}$, i.e., Eqs. (\ref{eq:ffAC}) and (\ref{eq:xAC}), in Eq. (\ref{eq: ff-xx0}), the empirical formula for the $f$-mode frequency (Eq.~(20) in Ref.~\cite{Sotani20b}) is updated as a function of $x$ and $\eta$ as 
\begin{gather}
    f_f(x,\eta)\ {\rm (kHz)}  = 1.6970x + f_0(\eta), \label{eq: ff-xx1} \\
    f_0(\eta)\ {\rm (kHz)}  = 0.53699\eta_{100}^{-1} + 0.29886 + 0.16476\eta_{100}. \label{eq: ff0}
\end{gather}

%%%%%%%%%%%%%%%%%%%%%%%%%%%%%%%%%%%
% Figure A2
%%%%%%%%%%%%%%%%%%%%%%%%%%%%%%%%%%%
%%%%%%%%%%%%%%%%%%%%%%%%%%%%%%%%%%%

\begin{figure}[tbp]
\begin{center}
\includegraphics[scale=0.5]{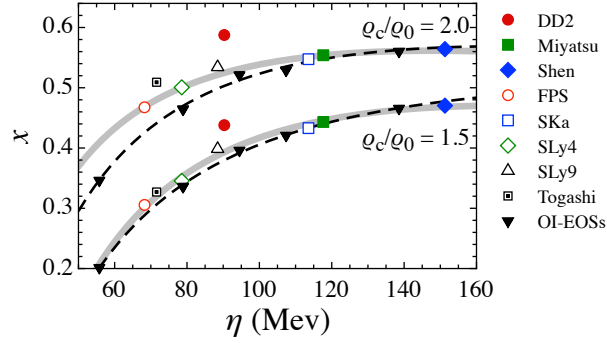}  
\end{center}
\caption{%%
The square root of the normalized stellar average density, $x$, for neutron star models constructed with various EOSs and with $\rho_c/\rho_0=1.5$ and 2, is shown as a function of $\eta$. The dashed lines correspond to the fitting lines derived in Ref. Ref. \cite{Sotani20b}, while the thick-solid lines are the fitting lines updated in this study with Eq. (\ref{eq:x-eta}).
}%%
\label{fig:x-eta}
\end{figure}

\begin{figure}[tbp]
\begin{center}
\includegraphics[scale=0.5]{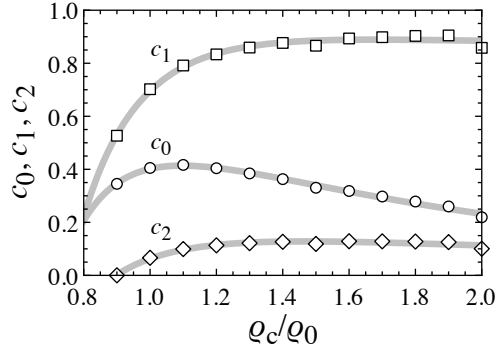}  
\end{center}
\caption{%%
The coefficients in Eq. (\ref{eq:x-eta}), $c_0$, $c_1$, and $c_2$, are shown as a function of $u_c=\rho_c/\rho_0$, where the thick-solid lines denote the fitting formulae given by Eqs. (\ref{eq:c0}) - (\ref{eq:c2}).
}%%
\label{fig:ci-uc}
\end{figure}

Furthermore, for low-mass neutron stars, the mass and the gravitational redshift are written as a function of $u_c$ and $\eta$, i.e., Eqs. (2) and (3) in Ref.~\cite{SIOO14}. In a similar way, we have shown that $x$ is also expressed as a function of $\eta$ and $u_c$ \cite{Sotani20b}. Here, we update the formula for $x$ with the EOSs considered in this study. In Fig.~\ref{fig:x-eta}, we show the values of $x$ for the neutron star models with $\rho_c/\rho_0=2.0$ and 1.5 for various EOSs as a function of $\eta$. With these data together with the data obtained with the OI-EOSs in Ref.~\cite{Sotani20b} except for the case with $\eta=55.8$ MeV, we can find that $x$ is fitted well with the same function form as in Ref.~\cite{Sotani20b}, such as
\begin{equation}
  x = -c_0\eta_{100}^{-1} + c_1 - c_2\eta_{100}, \label{eq:x-eta}
\end{equation}
where $c_0$, $c_1$, and $c_2$ are positive coefficients depending on $u_c=\rho_c/\rho_0$. In addition, as in Fig.~\ref{fig:ci-uc}, these coefficients are fitted as a function of $u_c$, i.e.,
\begin{gather}
  c_0(u_c) = -0.9874u_c^{-3} + 1.2209u_c^{-2} + 0.2354u_c^{-1} - 0.067204, \label{eq:c0} \\
  c_1(u_c) = -1.4814u_c^{-3} + 2.0594u_c^{-2} - 0.8753u_c^{-1} + 0.9928,  \label{eq:c1} \\
  c_2(u_c) = -0.3760u_c^{-3} + 0.2481u_c^{-2} + 0.1859u_c^{-1} + 0.0050934. \label{eq:c2}
\end{gather}
We remark that we newly add the term of $u_c^{-3}$ in this study for fitting of the coefficients $c_i$. Now, we can get the square root of the normalized stellar average density as a function of $\eta$ and $u_c$, i.e., $x=x(\eta,u_c)$ with Eqs. (\ref{eq:x-eta}) - (\ref{eq:c2}). So, the empirical formula for the $f$-mode frequency, i.e., Eq. (\ref{eq: ff-xx1}), is written as a function of $\eta$ and $u_c$ instead of a function of $\eta$ and $x$, for low-mass neutron stars, i.e., $f_f=f_f(u_c,\eta)$. This is the updated version with using the realistic EOSs considered in this study. Together with the mass formula, i.e., $M=M(u_c,\eta)$, given by Eq. (2) in Ref.~\cite{SIOO14}, one would estimate the values of $\eta$ and $u_c$ via the observation of the $f$-mode frequency from the neutron star whose mass is known.

%%%%%%%%%%%%%%%%%%%%%%%%%%%%%%%%%%%
% Figure A3
%%%%%%%%%%%%%%%%%%%%%%%%%%%%%%%%%%%
%%%%%%%%%%%%%%%%%%%%%%%%%%%%%%%%%%%

%%%%%%%%%%%%%%%%%%%%%%%%%%%%%%%%%%%
% Figure A4
%%%%%%%%%%%%%%%%%%%%%%%%%%%%%%%%%%%
%%%%%%%%%%%%%%%%%%%%%%%%%%%%%%%%%%%

\begin{figure*}[tbp]
\begin{center}
\includegraphics[scale=0.45]{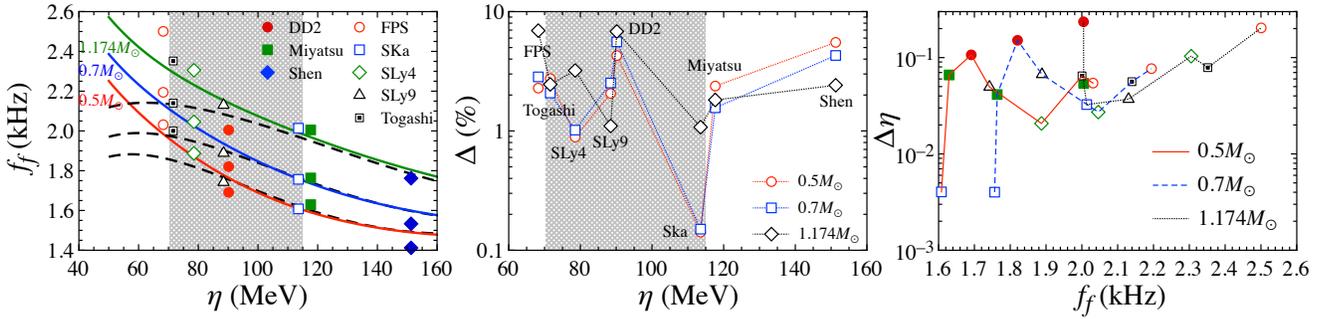} 
\end{center}
\caption{%%
In the left panel, the $f$-mode frequencies estimated with the empirical formula from the neutron star model with $M/M_\odot=0.5$, 0.7, and 1.174 are shown with the solid line as a function of $\eta$, where the lines from bottom up correspond to the results for $M/M_\odot=0.5$, 0.7, and 1.174. The marks denote the frequency calculated for each neutron star model. For reference, the $f$-mode frequencies estimated with the empirical formula derived in Ref.~\cite{Sotani20b} are also shown with dashed lines. In the middle panel, the relative deviation of the $f$-mode frequencies estimated with empirical formula from that calculated for each neutron star model is shown as a function of $\eta$, where the circle, square, and diamond correspond to the results for $M/M_\odot=0.5$, 0.7, and 1.174. In the both panels, the shaded region denotes the fiducial value of $\eta$ constrained by terrestrial experiments. In the right panel, the relative deviation of $\eta$ estimated with the empirical formulae for the mass and the $f$-mode frequency from the value of $\eta$ for each stellar model is shown as a function of the $f$-mode frequency, where the solid, dashed, and dotted lines correspond to the neutron star models with $M/M_\odot=0.5$, 0.7, and 1.174, respectively.
}%%
\label{fig:Delta}
\end{figure*}

Finally, we check how well the $f$-mode frequency from the low-mass neutron star whose mass is known, is estimated with the empirical formulae of the $f$-mode frequency and mass. For this purpose, we consider the neutron star models with $M/M_\odot=0.5$, 0.7, and 1.174. We remark that the minimum mass of neutron star observed so far is $M=1.174M_\odot$ \cite{Martinez15}. Once the mass would be fixed (or observed), one can derive the relation between $\eta$ and $u_c$ via the mass formula. With this relation, one can estimate the $f$-mode frequency as a function of $\eta$ (or $u_c$), using the empirical formula for the $f$-mode frequency. In the left panel of Fig.~\ref{fig:Delta}, we show the $f$-mode frequency estimated in such a way with the solid lines, where the lines from bottom up correspond to the neutron star models with $M/M_\odot=0.5$, 0.7, and 1.174. Meanwhile, the marks in this figure denote the $f$-mode frequencies calculated for each neutron star model. In addition, the dashed lines denote the estimation of the $f$-mode frequency with the empirical formula derived in Ref.~\cite{Sotani20b}. It is obviously  found that the empirical formula updated in this study works well even in the region with lower value of $\eta$, unlike the original empirical formula. Moreover, in order to estimate the deviation of the frequency estimated with the empirical formulae from that calculated for each neutron star model, we calculate the relative deviation defined by
\begin{equation}
  \Delta = \frac{|f_f - f_f(u_c,\eta)|}{f_f}, \label{eq:Delta}
\end{equation}
where $f_f$ is the $f$-mode frequencies calculated for each stellar model, while $f_f(u_c,\eta)$ denotes the frequency estimated with the empirical formula given by Eqs. (\ref{eq: ff-xx1}) - (\ref{eq:c2}) together with the mass formula, i.e., $M=M(u_c,\eta)$, given by Eq. (2) in Ref.~\cite{SIOO14}. We remark that the relative deviation discussed here is different from the relative deviation given by Eq. (\ref{eq:abs_dff}), where $f_f(x,\eta)$ in Eq. (\ref{eq:abs_dff}) is calculated via Eq. (\ref{eq:ff_x}) by using the value of $x$ and $\eta$ for each neutron star model. 
%We also remark that the relative deviation given by Eq. (\ref{eq:Delta}) is different from that given by Eq. (\ref{eq:Delta_GW}), i.e., $f_f^{\rm (em;M)}$ in Eq. (\ref{eq:Delta}) is the $f$-mode frequency estimated with the mass formula and the empirical formula for the $f$-mode for low-mass neutron stars as a function of $u_c$ and $\eta$, while $f_f^{\rm (em)}$ in Eq. (\ref{eq:Delta_GW}) is that estimated with the empirical formulae for the $f$- and $p_1$-modes as a function of $x$ and $\eta$. 
The resultant values of the relative deviation with Eq. (\ref{eq:Delta}) are shown in the middle panel of Fig.~\ref{fig:Delta}, where the circle, square, and diamond correspond to the results for the neutron star models with $M/M_\odot=0.5$, 0.7, and 1.174. From this figure, we find that the $f$-mode frequency from not only the neutron star model whose mass is less than $\sim 0.7M_\odot$ but also that with $M\sim M_\odot$ can be estimated well via the empirical formulae. 
Furthermore, in a similar way to the right panel of Fig. \ref{fig:Delta-eta}, we show the relative deviation of $\eta$ estimated with the empirical formulae for the mass and the $f$-mode frequency from the value of $\eta$ for each stellar model in the right panel of Fig. \ref{fig:Delta}, where the solid, dashed, and dotted lined correspond to the neutron star models with $M/M_\odot=0.5$, 0.7, and 1.174, respectively. Here, the relative deviation, $\Delta \eta$, is similarly calculated by Eq. (\ref{eq:Delta_eta}), but $\eta(f_f,f_{p_1})$ in the equation is replaced by $\eta(M,f_f)$, which is the value of $\eta$ estimated with the empirical formulae for the mass and the $f$-mode frequency. From this figure, we find that the value of $\eta$ can be estimated within $\sim 20\%$ accuracy, once the $f$-mode frequency will be observed from the neutron star whose mass is known, if the stellar mass is less than $1.174M_\odot$.

%%%%%%%%%%%%%%%%%%%%%%%%%%%%%%%%%%%
% Figure A5
%%%%%%%%%%%%%%%%%%%%%%%%%%%%%%%%%%%
%%%%%%%%%%%%%%%%%%%%%%%%%%%%%%%%%%%

%\bibliographystyle{h-physrev} % for PrD
%\bibliography{mybib}

\begin{thebibliography}{999}
\bibitem{note_DD2}  % NOTE:DD2
   According to the original study for DD2 \cite{DD2a}, the value of $L$ is found to be $57.94$ MeV, which leads to $\eta=93.4$ MeV. With this value, the deviation of the plot for DD2 from the fitting line in Figs. \ref{fig:Delta-eta}, \ref{fig:fAC-eta}, \ref{fig:x-eta}, and \ref{fig:Delta} might slightly become small.
\end{thebibliography}
%%%%%%%%%%%%%%%%%%%%%%%%%%%%%%%%%%%%%%%%%%%%%%%%

\end{document}